\RequirePackage{ifpdf}
\documentclass[paper]{JHEP3}
\usepackage{amssymb,enumerate}
\usepackage{amsmath}
\usepackage{bbm}
\usepackage{url}
\usepackage{cite}
\usepackage{graphics}
\usepackage{xspace}
\usepackage{epsfig}
\usepackage{subfigure}
\usepackage{booktabs}

\newcommand\POWHEG{{\tt POWHEG}}
\newcommand\POWHEGBOX{{\tt POWHEG~BOX}}

\newcommand\PYTHIA{{\tt PYTHIA}}

\newcommand\HERWIGPP{{\tt HERWIG++}}
\newcommand\HERPP{{\tt HER++}}
\newcommand\DSPP{{\tt DS++}}
\newcommand\PYT{{\tt PYT}}

\def\beq{\begin{equation}}
\def\eeq{\end{equation}}

\def\mr{\mathrm}
\def\vbfh3j{VBF $Hjjj$\;}
\def\mc{\mathcal}
\def\muf{\mu_\mr{F}}
\def\mur{\mu_\mr{R}}

\title{Parton-shower effects on Higgs boson production via vector-boson fusion in association with three jets} 
\vfill
\author{
  Barbara J\"ager \\
  Institute for Theoretical Physics, University of T\"ubingen,  Auf der Morgenstelle 14, 72076 T\"ubingen, Germany\\
  E-mail: \email{jaeger@itp.uni-tuebingen.de} }

\author{Franziska Schissler \\
  Institute for Theoretical Physics, Karlsruhe Institute of Technology, 76128 Karlsruhe, Germany\\
  E-mail: \email{franziska.schissler@kit.edu} }

\author{Dieter Zeppenfeld \\
 Institute for Theoretical Physics, Karlsruhe Institute of Technology, 76128 Karlsruhe, Germany \\
  E-mail: \email{dieter.zeppenfeld@kit.edu} }


\abstract{
We present an implementation of Higgs boson production via vector-boson fusion in association with three jets at hadron colliders in the \POWHEGBOX{}, a framework for the matching of NLO-QCD calculations with parton-shower programs. 
Our work provides the means to precisely describe the properties of extra jet activity in vector-boson fusion reactions that are used for the suppression of QCD backgrounds by central jet veto techniques. 
For a representative setup at the CERN LHC we verify that uncertainties related to parton-shower effects are mild for distributions related to the third jet, in contrast to what has been observed in calculations based on vector-boson fusion induced Higgs production in association with two jets. 
}

\begin{document}

\section{Introduction}
With the discovery of the Higgs boson at the CERN Large Hadron Collider (LHC) \cite{atlas:2012gk,cms:2012gu} particle physics has entered a new era. Both LHC collaborations, ATLAS and CMS, have confirmed the existence of a boson with a mass of about 126~GeV and properties consistent with those of the scalar CP-even particle predicted by the Standard Model \cite{Aad:2013xqa,Chatrchyan:2012jja}. In order to fully establish the nature of the Higgs boson, a precise determination of its couplings to fermions and gauge bosons is essential \cite{Zeppenfeld:2000td,Duhrssen:2004cv,LHCHiggsCrossSectionWorkingGroup:2012nn}.  

A rather clean environment for such coupling measurements is provided by the vector-boson fusion (VBF) production mode \cite{Rainwater:1998kj,Plehn:1999xi,Rainwater:1999sd,Kauer:2000hi,Rainwater:1997dg,Eboli:2000ze}, where the Higgs boson is produced via quark-scattering mediated by weak gauge boson exchange in the $t$-channel, $qq'\to qq'H$. Because of the low virtuality of the exchanged weak bosons, the tagging jets emerging from the scattered quarks are typically located in the forward and backward regions of the detector, while the central-rapidity region exhibits little jet activity due to the color-singlet nature of the $t$-channel exchange. These features can be exploited to efficiently suppress QCD backgrounds with a priori large cross sections at the LHC. 
In the context of central-jet veto (CJV) techniques, events are discarded if they exhibit one or more jets in between the two tagging jets. To quantitatively employ such selection strategies, a precise knowledge of the VBF cross section with an additional jet, i.e. the reaction $pp\to H jjj$, is crucial. 

Next-to-leading order (NLO) QCD corrections to VBF-induced $Hjjj$ production have first been computed in \cite{Figy:2007kv}, yielding results with only small residual scale uncertainties of order 10\% or less. In particular, in that approach the survival probability for the Higgs signal has been estimated to exhibit a perturbative accuracy of about 1\%. The calculation of \cite{Figy:2007kv} is implemented in the {\tt VBFNLO} package~\cite{Arnold:2008rz,Arnold:2011wj,Arnold:2012xn} in the form of a flexible parton-level Monte-Carlo program.   
More recently, an NLO-QCD calculation for electroweak $Hjjj$ production has been presented~\cite{Campanario:2013fsa}, where several approximations of Ref.~\cite{Figy:2007kv} have been dropped. 

In this work, we merge the parton-level calculation of \cite{Figy:2007kv} with a parton-shower Monte-Carlo in the framework of the \POWHEG{} formalism~\cite{Nason:2004rx,Frixione:2007vw}, a method for the matching of an NLO-QCD calculation with a transverse-momentum ordered parton-shower program. For our implementation we are making use of version~2 of the  \POWHEGBOX{}~\cite{Alioli:2010xd,Nason:2013ydw}, a repository that provides the process-independent ingredients of the \POWHEG{} method. The code we develop yields precise, yet realistic predictions for VBF-induced $Hjjj$ production at the LHC in a public framework that can easily be used by the reader for further phenomenological studies. 

This article is organized as follows: In Sec.~\ref{sec:tech} we describe some technical details of our implementation. Phenomenological results are presented in Sec.~\ref{sec:pheno}. We conclude in Sec.~\ref{sec:concl}.
%
%
\section{Technical details of the implementation}
\label{sec:tech}
The implementation of $Hjjj$ production via VBF in the context of the \POWHEGBOX{} requires, as major building blocks, the matrix elements for all relevant partonic scattering processes at Born level and at next-to-leading order. These have first been calculated in \cite{Figy:2007kv} and are publicly available in the {\tt VBFNLO} package~\cite{Arnold:2008rz}. We extracted the matrix elements from {\tt VBFNLO} and adapted them to the format required by the \POWHEGBOX. 

\begin{figure}[t]
 \centering
 \includegraphics[width=\textwidth]{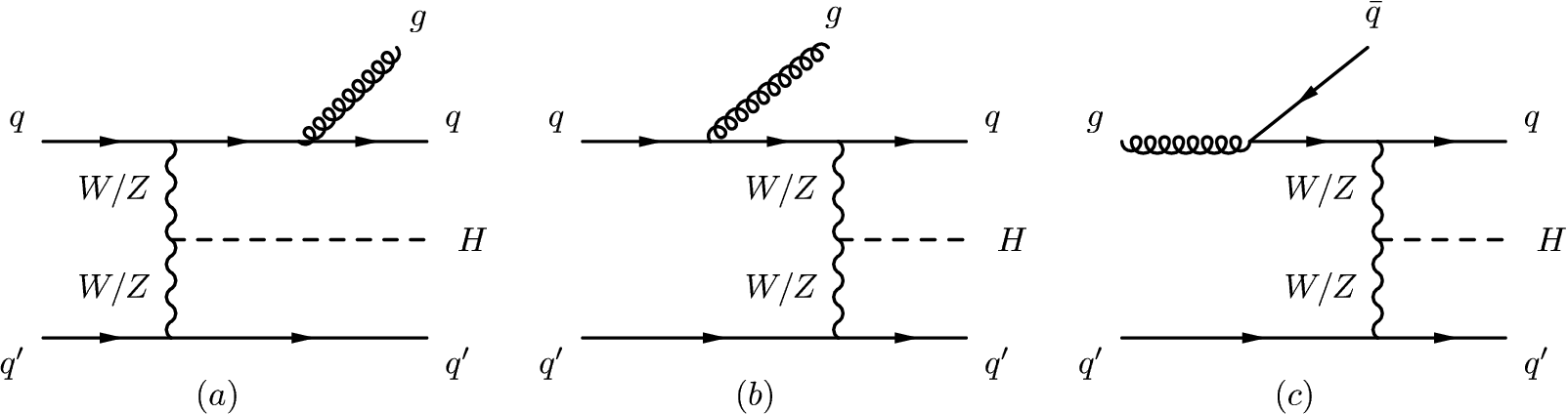}
 \caption{Representative tree-level diagrams for VBF $Hjjj$ production. }
 \label{fig:lo-graphs}
\end{figure}

At leading order (LO), processes of the type $qq'\to qq'gH$ and all crossing-related channels are taken into account, if they include the exchange of a weak boson in the $t$-channel . Some representative Feynman diagrams are depicted in Fig.~\ref{fig:lo-graphs}. 
The gauge-invariant class of diagrams involving weak-boson exchange in the $s$-channel is considered as part of the Higgs-strahlung process, and disregarded in the context of our work on VBF-induced Higgs production. The interference of $t$-channel with $u$-channel diagrams in flavor channels with quarks of the same type is neglected. Once VBF-specific selection cuts are imposed, these approximations are well justified~\cite{Ciccolini:2007ec}.
Throughout, we assume a diagonal CKM matrix. 
We refer to the electroweak $Hjjj$ production process at order $\mc{O}(\alpha_s\alpha^3)$ within these approximations as ``VBF $Hjjj$ production''.  

\begin{figure}[tp]
\vspace*{8cm}
 \includegraphics[width=\textwidth,bb=0 0 447 0]{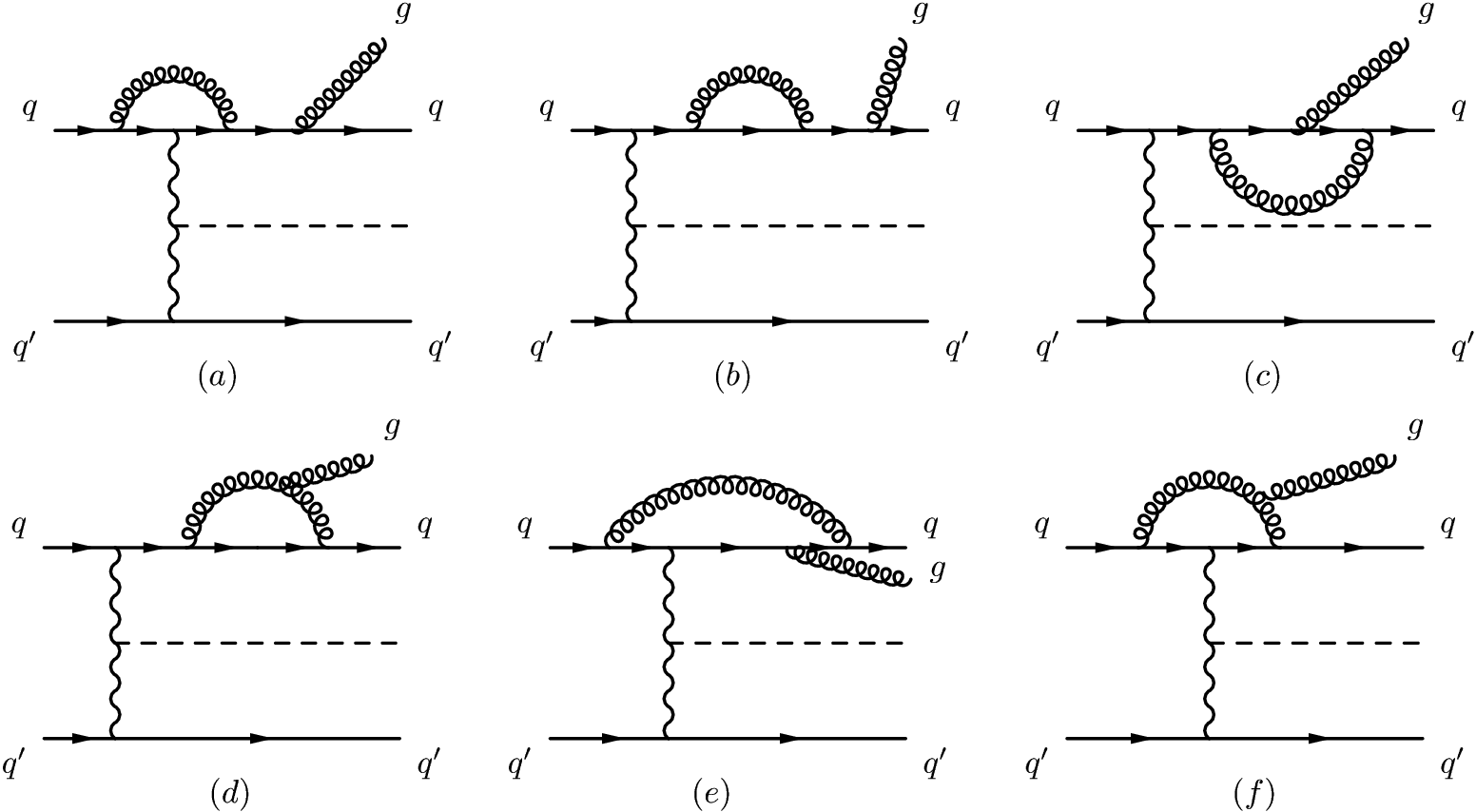}
 \\
 \includegraphics[width=\textwidth]{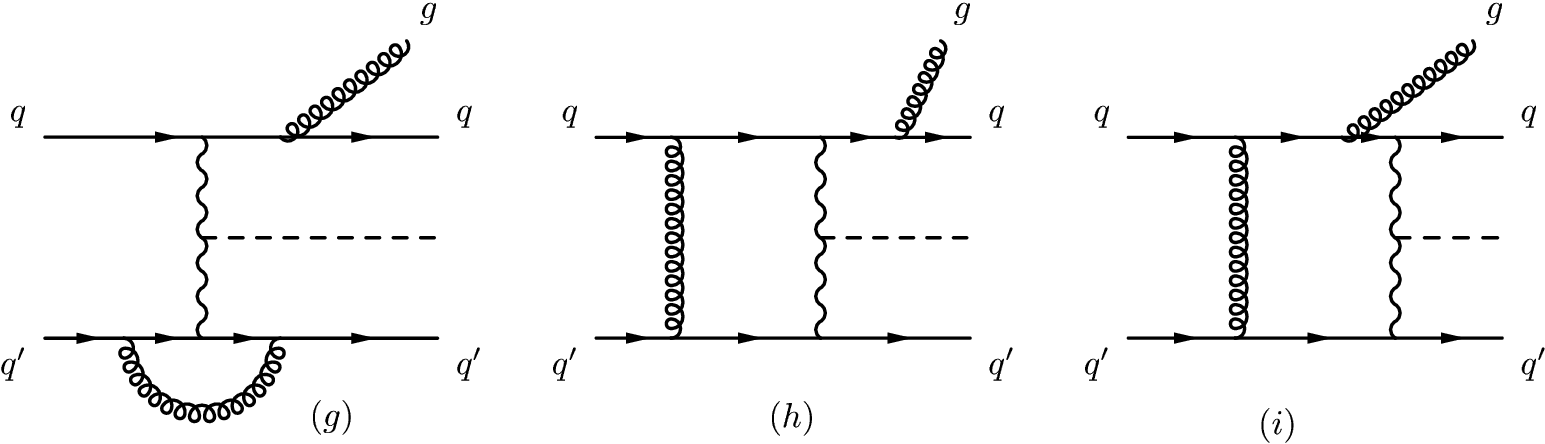}
 \caption{
 Representative one-loop diagrams for $qq'\to qq'gH$, with the virtual gluon being attached to one single fermion line [graphs ($a$)-($g$)], or to the two different fermion lines  [graphs ($h$) and ($i$)]. }
 \label{fig:virt-graphs}
\end{figure}

The virtual corrections to this reaction comprise the interference of the Born amplitudes with one-loop diagrams where a virtual gluon is attached to a single fermion line [c.f.~Fig.~\ref{fig:virt-graphs}~(a)--(g)], and diagrams where a virtual gluon is exchanged between the two different fermion lines, see~Fig.~\ref{fig:virt-graphs}~(h),(i). 
As discussed in some detail in \cite{Figy:2007kv}, the 
latter contributions are strongly suppressed by color factors and due to the VBF dynamics. They can be neglected, if the respective color structures of the real-emission contributions are disregarded as well, as these would serve to cancel the  infrared singularities of the pentagon and hexagon contributions that we drop. 

\begin{figure}[tp]
 \includegraphics[width=1\textwidth]{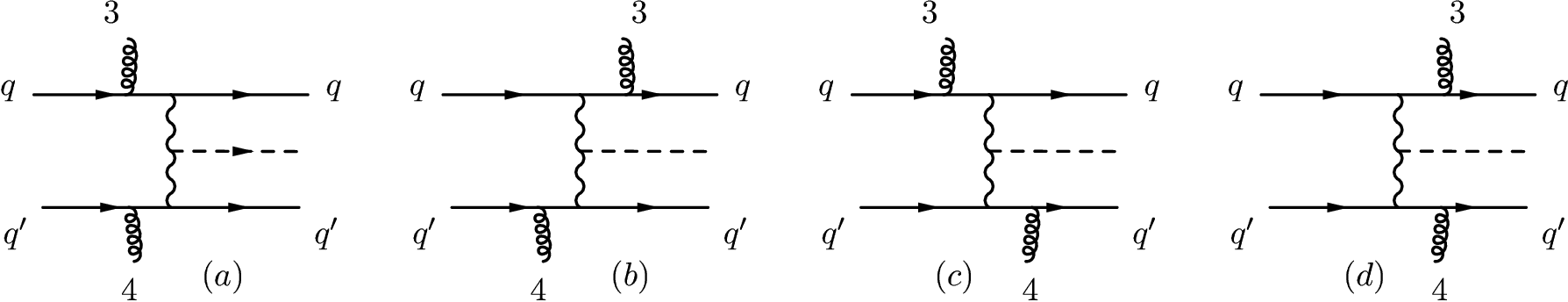}
 \caption{
Representative diagrams of the color structure $\mc{B}_4^3$ as introduced in Ref.~\cite{Figy:2007kv} for the subprocess $qq' \rightarrow  qq'\, gg \, H$. 
}
 \label{fig:real-gg-supp}
\end{figure}

\begin{figure}[tp]
\begin{center}
 \includegraphics[width=1\textwidth]{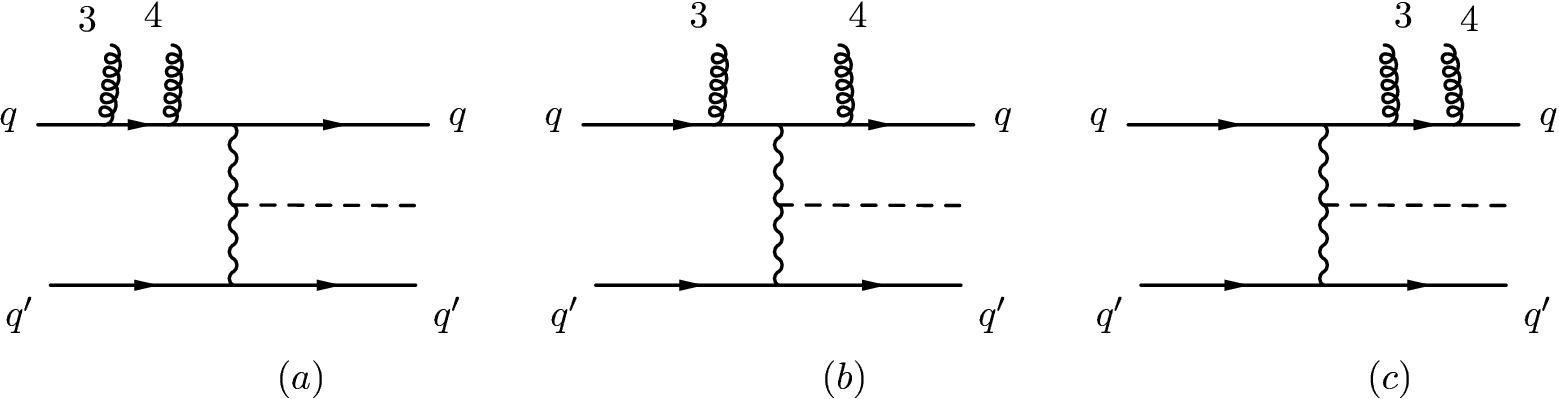} \\
 \vspace{0.5cm} 
 \hspace{1cm} \includegraphics[width=0.66\textwidth]{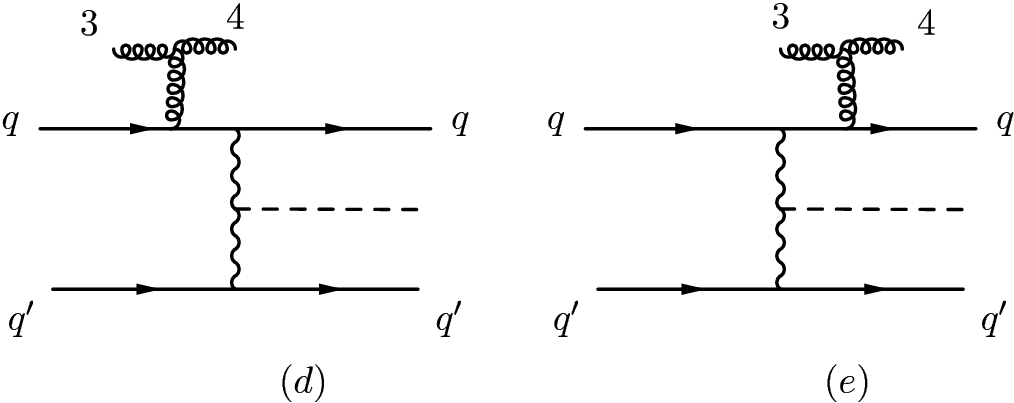}  
 \caption{Representative diagrams  of the color structure $\mc{A}_{43}^{1a}$ as introduced in Ref.~\cite{Figy:2007kv} for the subprocess $qq' \rightarrow  qq'\, gg \, H$. }
 \label{fig:real-gg}
 \end{center}
\end{figure}

\begin{figure}[tp]
 \centering
 \includegraphics[width=1\textwidth]{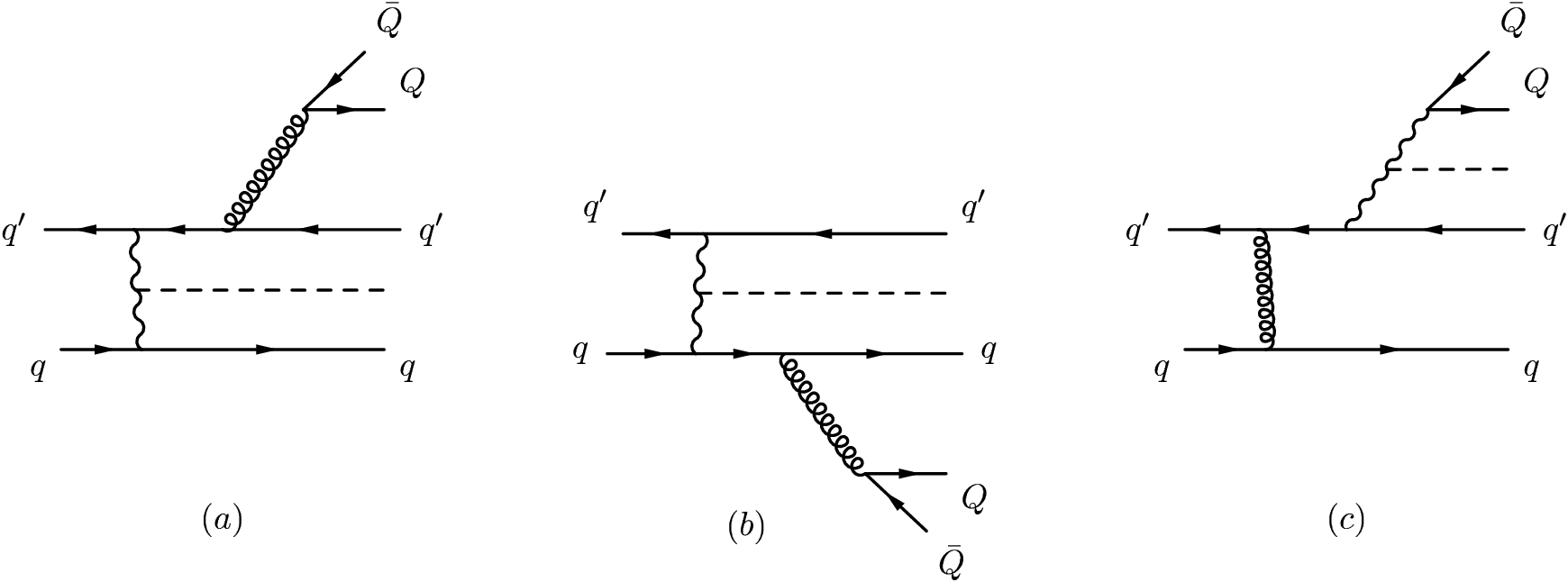}
 \caption{
 Representative diagrams for the subprocess $qq' \to qq' \, Q\bar{Q}\, H$. }
 \label{fig:real-QQ}
\end{figure}

The real-emission contributions involve subprocesses with four external (anti-)quarks and two gluons such as $qq'\to qq'ggH $, as well as pure quark scattering processes of the type $qq'\to qq'Q\bar QH $, and all crossing-related channels with $t$-channel weak boson exchange. 
Because of the approximations we have employed in the virtual contributions, where we dropped color-suppressed contributions giving rise to pentagon and hexagon integrals, the respective color structures have to be disregarded in the real-emission contributions as well. In practice this means to neglect interference terms between diagrams where a given gluon is emitted once from the upper and once from the lower quark line in the Feynman graphs. For example, interference terms like $2\,\mc{R}e\left( {\cal B}^3_4 {\cal B^*}^4_3\right)$, with ${\cal B}^3_4$ as depicted in Fig.~\ref{fig:real-gg-supp}, are dropped, 
while their squares or the squares of the topologies sketched in Fig.~\ref{fig:real-gg} are fully considered. 
Representative diagrams for a pure quark subprocess are depicted in Fig.~\ref{fig:real-QQ}.  For this class of subprocesses we require that the $Q\bar Q$ pair stems from a gluon. Contributions involving the hadronic decay of a weak boson, $V\to Q\bar Q$, such as graph~\ref{fig:real-QQ}~(c),  are disregarded within our VBF~setup. 

%
While in \cite{Figy:2007kv} soft and collinear singularities have been taken care of by a dipole subtraction procedure, the \POWHEGBOX{} makes use of the so-called FKS subtraction scheme \cite{Frixione:1995ms}. From the color- and spin-correlated amplitudes provided by the user, the \POWHEGBOX{} internally constructs the counterterms that are needed to cancel soft and collinear singularities in the real-emission contributions. Because we are disregarding certain color-suppressed contributions in the virtual and real-emission amplitudes, we have to make sure that only the counterterms relevant for our setup are constructed. This is achieved by passing only those  color- and spin-correlated Born amplitudes to the \POWHEGBOX{} that correspond to the color structures we consider within our approximations. 

We have carefully tested that the counter terms constructed in this way approach the real-emission amplitudes in the soft and collinear limits. Additionally, we have compared the tree-level and real emission amplitudes for selected phase space points with code generated by {\tt MadGraph} \cite{Stelzer:1994ta} that has been adapted to match the approximations of our calculation. We found agreement at the level of more than ten digits. The virtual amplitudes have been compared to {\tt VBFNLO}, again showing full agreement for single phase space points. We note that some care has been necessary in this latter check, as finite parts of the subtraction terms are included in the virtual amplitudes in the default setup of {\tt VBFNLO}.  
To verify the entire setup of our code, we have compared cross sections and distributions for various sets of selection cuts at LO and NLO-QCD accuracy as obtained with the \POWHEGBOX{} with respective results of {\tt VBFNLO}. We found full agreement for all considered scenarios. 

We note that special care is needed when performing the phase-space integration of \vbfh3j~production in the framework of the \POWHEGBOX{}. In contrast to the VBF-induced $Hjj$ production cross section that is entirely finite at leading order, the inclusive VBF $Hjjj$ cross section diverges already at leading order when a pair of partons becomes collinear  or a soft gluon is encountered in the final state. While such divergent contributions disappear after phenomenologically sensible selection cuts are imposed, their presence considerably reduces the efficiency of the numerical phase space integration. This effect can be avoided by appropriate phase-space cuts at generation level, or by a so-called Born-suppression factor $F(\Phi_n)$ that dampens the integrand whenever singular configurations in phase-space are approached. In order to 
ensure that our results are independent of technical cuts in the phase-space integration, we recommend the use of a Born-suppression factor. In our  \POWHEGBOX{} implementation we provide two alternative versions of Born-suppression factors: 
\begin{itemize}
\item
In our first, multiplicative, approach, the factor is of the form 
\begin{equation}
\label{eq:bsupp1}
F(\Phi_n) = 
\prod_{i=1}^3
\left(
\frac{p_{T,i}^2}{p_{T,i}^2+\Lambda_p^2}
\right)^2
\prod_{i,j=1;\\
j\neq i}^3
\left(
\frac{m_{ij}^2}{m_{ij}^2+\Lambda_m^2}
\right)^2\,,
\end{equation}
where the $p_{T,i}$ and $m_{ij}=\sqrt{(p_i+p_j)^2}$ respectively denote the transverse momenta and invariant masses of the three final-state partons of the underlying Born configuration. The $\Lambda_p$ and $\Lambda_m$ are cutoff parameters that are typically set to values of a few GeV.   %
\item
Following the procedure suggested for the related case of trijet production in the framework of the \POWHEGBOX{}~\cite{Kardos:2014dua}, we use an exponential suppression factor of the form
\begin{equation}
S_1 = \exp \left[ 
- \Lambda_1^4 \cdot \left(\sum_{i=1}^3 \frac{1}{p^4_{T,i}} + \sum_{i,j=1;\\
j\neq i}^3 \frac{1}{q^2_{ij}}\right) 
\right]\,,
\end{equation}
with 
\beq
q_{ij}=p_i \cdot p_j \, \frac{E_i \, E_j}{E_i^2 + E_j^2}\,,
\eeq
for the suppression of infrared divergent configurations in the underlying Born kinematics, 
accompanied by a factor 
\begin{equation}
 S_2 = \left(\frac{H_T^2}{H_T^2 + \Lambda_2^2}\right)^2,
\end{equation}
where
\beq
H_T = p_{T,1} + p_{T,2} + p_{T,3}.
\eeq
The factor $S_2$ serves to suppress configurations where all partons are having small transverse momenta, and at the same time increase the fraction of events generated with large transverse momenta. 
Combining $S_1$ with $S_2$, we construct
\begin{equation}
\label{eq:bsupp2}
F(\Phi_n) = S_1 \cdot S_2\,.
\end{equation}
\end{itemize}
For the generation of the phenomenological results shown below we are using a Born suppression factor of the form given in Eq.~(\ref{eq:bsupp2}) with $\Lambda_1 = 10 $~GeV and  $\Lambda_2 = 30$~GeV, supplemented by a small generation cut on the transverse momenta of the three outgoing partons of the underlying Born configuration, $p_{T,i}^\mr{gen} > 1$~GeV.  

To make sure our results do not depend on these technical parameters, in addition to our default setup we ran our code using the Born suppression factor of Eq.~(\ref{eq:bsupp1}) with $\Lambda=20$~GeV and, again, $p_{T,i}^\mr{gen} > 1$~GeV. The results in the two setups are in full agreement with each other and, at fixed order, also with respective results obtained with {\tt VBFNLO} that is using an entirely different phase-space generator. 
%
%
\section{Phenomenological results}
\label{sec:pheno}
Our implementation of \vbfh3j production at the LHC is made publicly available in version~2 of  the \POWHEGBOX{}, and can be obtained as explained at the project webpage,  {\tt http://powhegbox.mib.infn.it/}. 

Here, we are providing phenomenological results for a representative setup at the LHC with a center-of-mass energy of $\sqrt{s}=8$~TeV. 
We are using the CT10 fixed-four-flavor set \cite{Lai:2010vv} for the parton distribution functions of the proton as implemented in the {\tt LHAPDF} library \cite{Whalley:2005nh} and the accompanying value of the strong coupling, $\alpha_s(m_Z)=0.1127$.  Jets are reconstructed via the anti-$k_T$ algorithm with a resolution parameter of $R=0.5$, with the help of the {\tt FASTJET}~package~\cite{Cacciari:2005hq,Cacciari:2008gp,Cacciari:2011ma}. As electroweak input parameters we are using the masses of the weak gauge bosons, $m_W=80.398$~GeV and $m_Z=91.188$~GeV, and the Fermi constant, $G_F=1.16639\times 10^{-5}$~GeV$^{-1}$. Other electroweak parameters are computed thereof via tree-level relations. 
The widths of the massive gauge bosons are set to $\Gamma_W =  2.095$~GeV and $\Gamma_Z=2.51$~GeV, respectively. For the Higgs boson, we are using $m_H = 126$~GeV and $\Gamma_H=4.095$~MeV. 
The renormalization and factorization scales are identified as $\mur=\muf=m_H/2$. 
In order to assess uncertainties that remain after matching the NLO calculation with a parton shower program, we consider three different tools:  \PYTHIA{}~{\tt 6.4.25} with the Perugia~0 tune~\cite{Sjostrand:2006za}, \HERWIGPP{}~{\tt 2.7.0}~\cite{Bahr:2008pv,Bellm:2013lba} with its default angular-ordered shower, and with a transverse-momentum ordered dipole shower~\cite{Platzer:2011bc} which we tag as {\tt PYT}, {\tt HER}, and {\tt DS++}, respectively. We note that wide-angle, soft radiation that is in principle needed when matching an NLO calculation with a parton-shower program using the \POWHEG{} method, is missing in the default angular-ordered \HERWIGPP{} shower. The impact of this missing piece on observables can only be estimated by a comparison with predictions obtained with transverse momentum ordered showers, such as the \DSPP{} version of \HERWIGPP{}. We do not consider hadronization, QED radiation, multiple parton interactions, and underlying event effects in this work. 

In order to define a $Hjjj$ event, we demand at least three well-observable jets with 
\beq 
p_{T,j}>20 \text{ GeV}\,,
\quad
|y_j| < 4.5\,.
\eeq
In addition, we impose VBF-specific selection cuts.  The two hardest jets, referred to as ``tagging jets'', are required to fulfill 
\beq
p_{T,j}^\mr{tag}>30~\mr{GeV}\,,\quad
|y_j^\mr{tag}|<4.5\,,
\eeq
and be well-separated from each other, 
\beq
|y_{j_1}^\mr{tag}-y_{j_2}^\mr{tag}|>4.0\,, 
\quad
y_{j_1}^\mr{tag}\times y_{j_2}^\mr{tag}<0\,, 
\quad
m_{jj}^\mr{tag}>500~\mr{GeV}\,.
\eeq
The kinematics of the Higgs boson is not restricted. 

\begin{figure}[t]
	\includegraphics[angle=-90,width=0.5\textwidth]{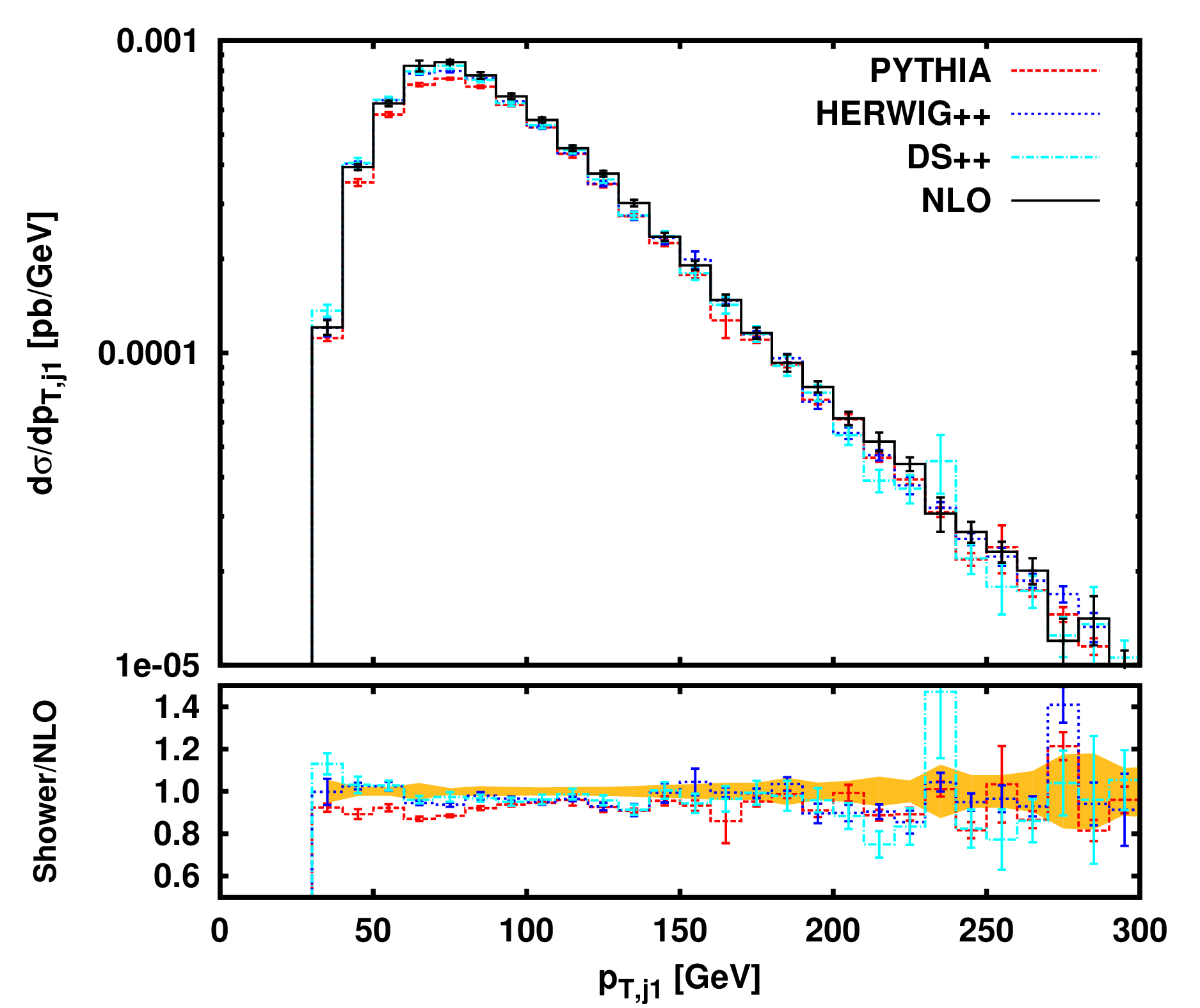}
	\includegraphics[angle=-90,width=0.5\textwidth]{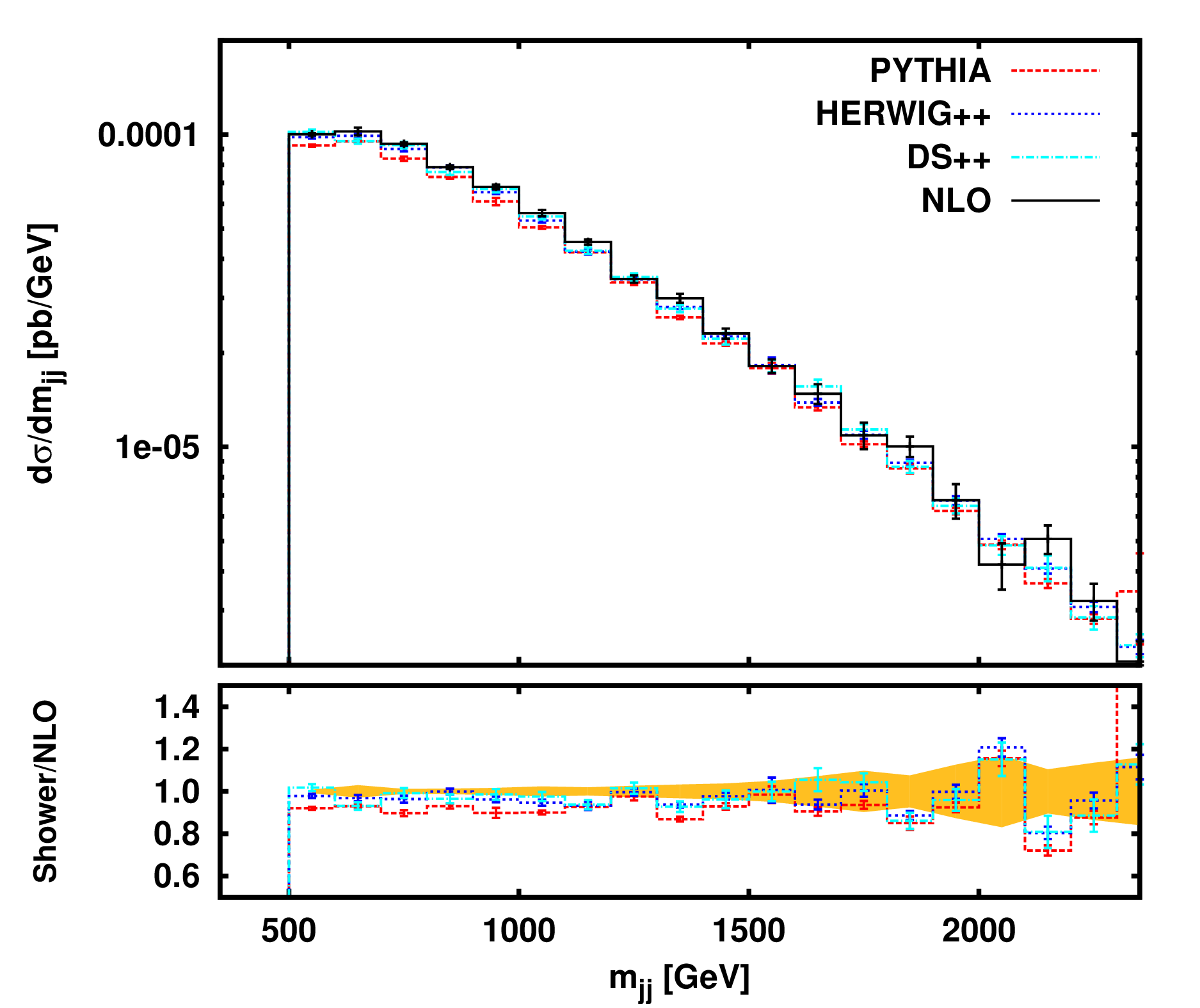}
\caption{
Transverse momentum distribution of the hardest tagging jet (left) and invariant mass distribution of the two tagging jets (right) at NLO (black), and at NLO+PS level: \PYT{}~(red), \HERPP{}~(blue), \DSPP{}~(cyan). 
The lower panels show the NLO+PS results normalized to the pure NLO prediction together with its statistical uncertainty (yellow band). 
}
\label{fig:tag-jet}
\end{figure}

With these cuts, we obtain a cross section of $\sigma^\mr{NLO}=71.5 \pm  0.4$~fb at fixed order, where the error is the statistical error of the Monte Carlo calculation. After matching the NLO result with a parton shower, some of the events fail to pass the cuts, resulting in slightly smaller cross sections of $\sigma^{\tt PYT}=65.8 \pm  0.3$~fb, $\sigma^{\tt HER}=68.3 \pm 0.3$~fb,  and $\sigma^{\tt DS++}=69.8\pm 0.5$~fb, respectively. 
Apart from this change in normalization the impact of the parton shower on observables related to the tagging jets is very mild, as  illustrated in Fig.~\ref{fig:tag-jet}
for the transverse momentum distribution of the hardest tagging jet and the invariant mass of the tagging jet pair. 

\begin{figure}[t]
	\includegraphics[angle=-90,width=0.5\textwidth]{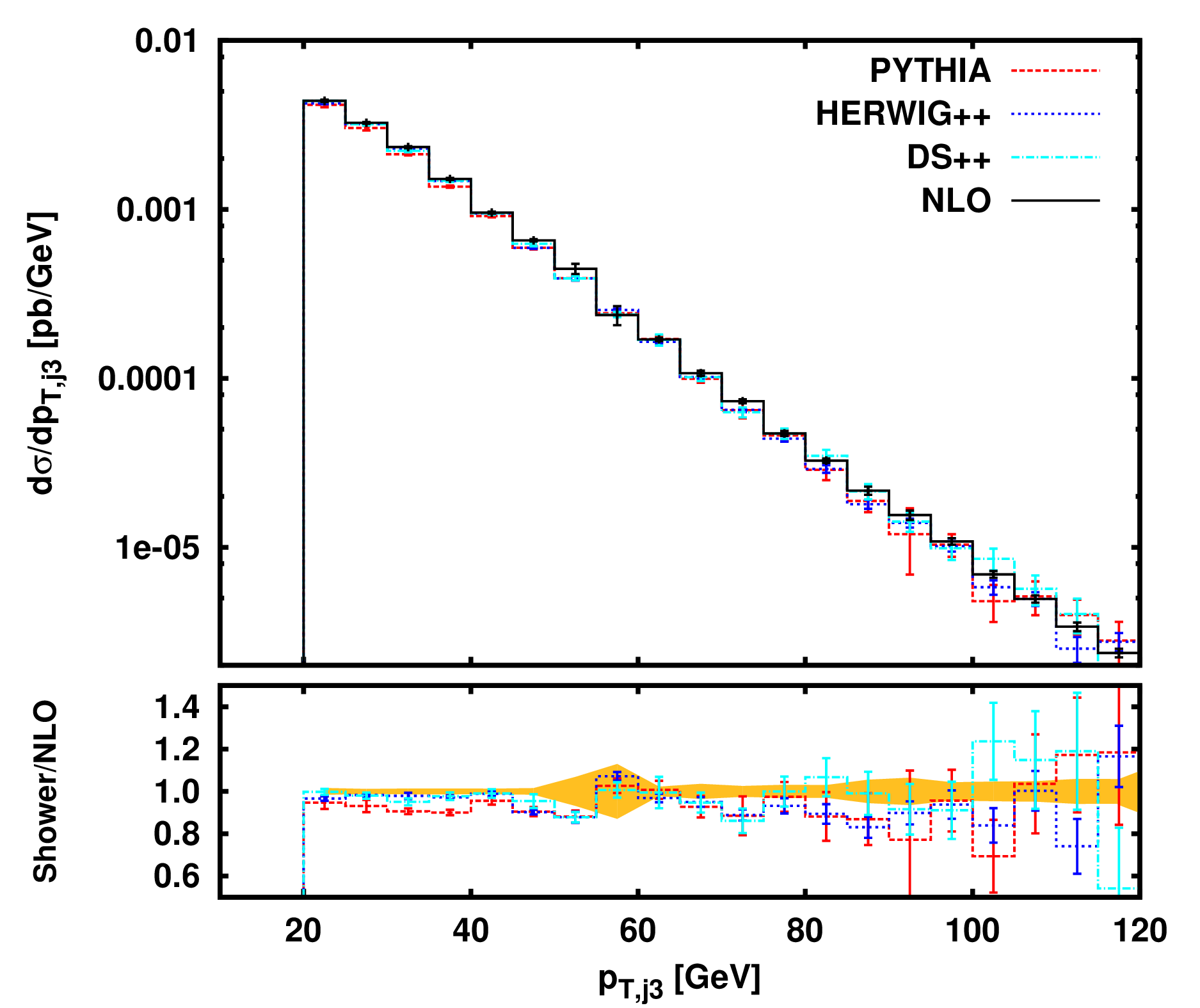}
	\includegraphics[angle=-90,width=0.5\textwidth]{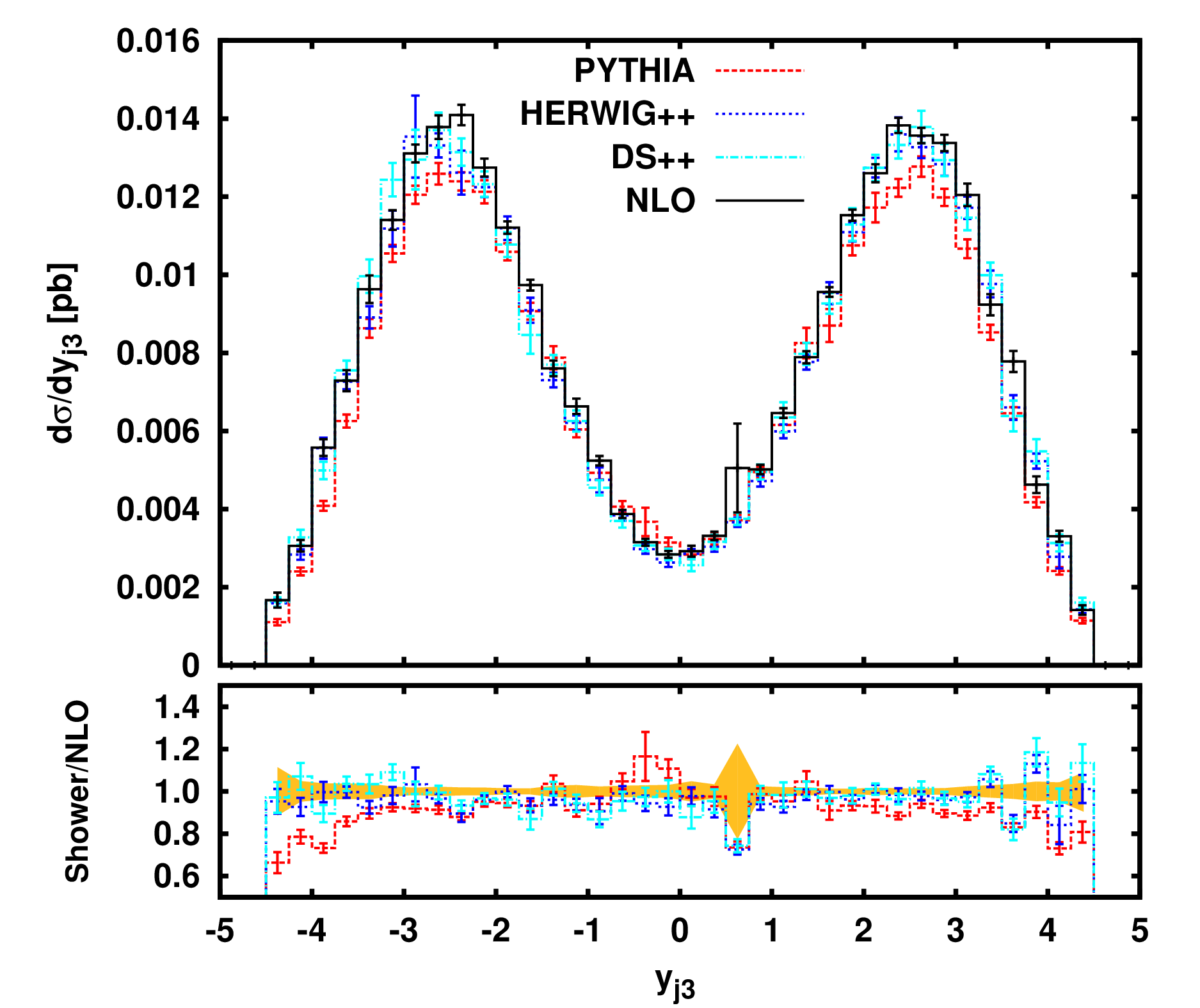}
\caption{
Transverse momentum and rapidity distributions of the third jet at NLO, and at NLO+PS level (line styles as in Fig.~\ref{fig:tag-jet}). 
}
\label{fig:jet3}
\end{figure}

In contrast to NLO calculations for VBF $Hjj$ production, where the third jet can be described only with LO accuracy, our calculation is NLO accurate in distributions related to the third jet. In Fig.~\ref{fig:jet3}, NLO+PS results for the transverse momentum and the rapidity distribution of the third jet are shown for different parton shower programs together with the fixed-order NLO result. 
For all considered parton showers, the difference between the NLO and the NLO+PS results is small. However, \PYTHIA{} tends to produce slightly more jets in the central-rapidity region of the detector, while \HERWIGPP{} preferentially radiates in the collinear region between the two tagging jets and the beam axis. We will see below that this effect is more pronounced in the case of sub-leading jets. 

\begin{figure}[t]
	\includegraphics[angle=-90,width=0.5\textwidth]{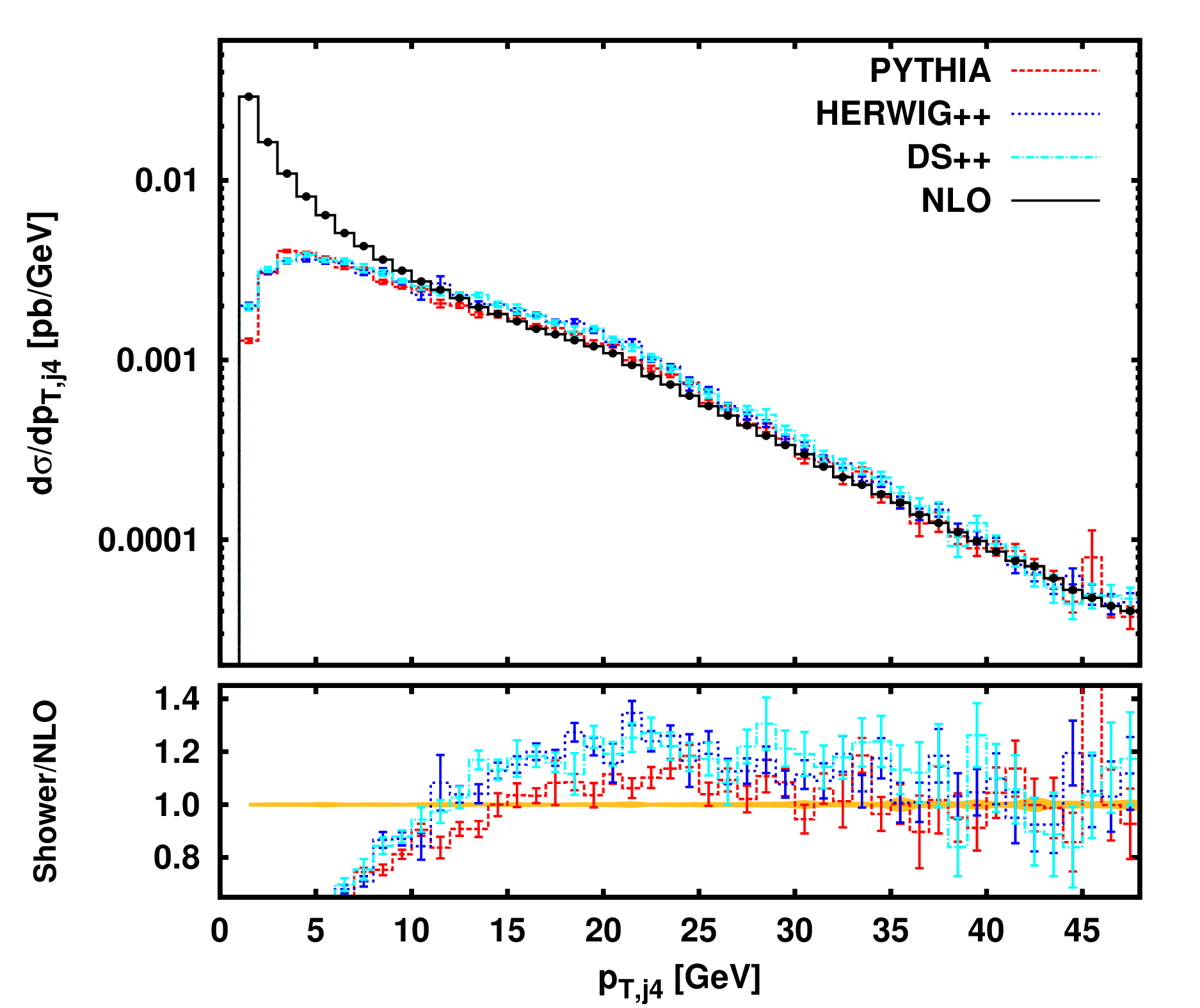}
	\includegraphics[angle=-90,width=0.5\textwidth]{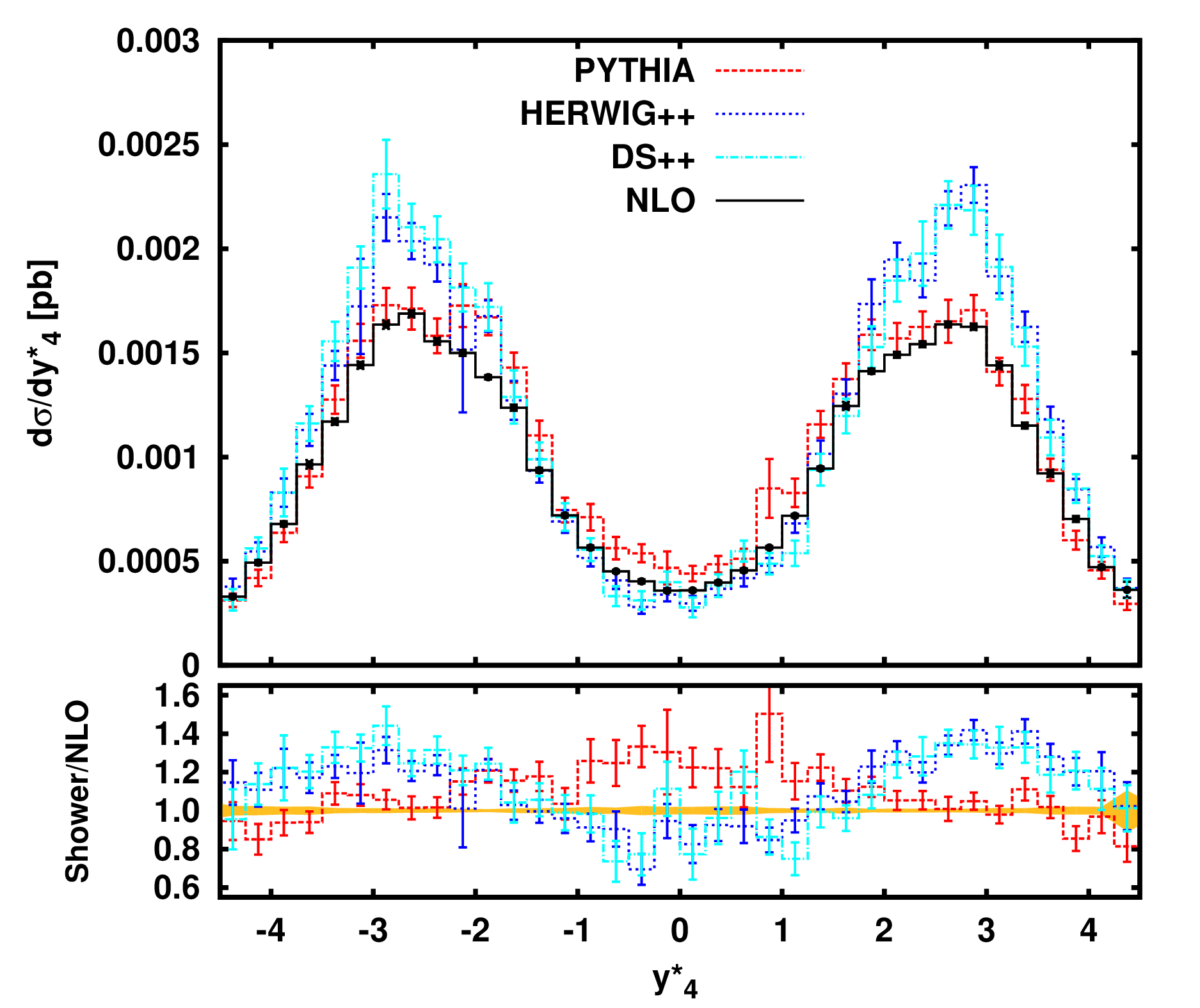}
\caption{
Transverse momentum distribution of a fourth jet for our default setup with an extra cut of $p_{T,j_4}>1$~GeV~(left) and rapidity distribution of a fourth hard jet with $p_{T,j_4}>20$~GeV relative to the two tagging jets (right) at NLO, and at NLO+PS level (line styles as in Fig.~\ref{fig:tag-jet}). 
}
\label{fig:jet4}
\end{figure}

Larger differences between the fixed-order and the various matched predictions occur in distributions related to the fourth jet. 
In the parton-level NLO calculation a fourth jet can only stem from the real-emission contributions, and can thus be described only at tree-level accuracy. Larger theoretical uncertainties are therefore expected for observables related to the fourth jet. Fig.~\ref{fig:jet4} 
illustrates the effect of the \POWHEG{}-Sudakov factor on the transverse momentum of the fourth jet and clarifies how extra radiation in the VBF setup is distributed by the different parton shower programs via the $y_4^\star$ variable. This quantity	 is defined as
\beq 
y_4^\star=y_{j_4} - \frac{y_{j_1} + y_{j_2}}{2}\,,
\label{eq:y4star}
\eeq
in order to parameterize the rapidity of the fourth jet relative to the two hard tagging jets. The respective distribution shows, more pronouncedly than in the case of the third jet, that \PYTHIA{} and \HERWIGPP{} tend to produce radiation in different regions of phase space. The differences between the various NLO+PS curves can thus be considered as inherent uncertainty of the matched prediction. 
%
%
\section{Conclusions}
\label{sec:concl}
In this work, we have presented an implementation of \vbfh3j production in version~2 of the \POWHEGBOX{} repository. We have performed the matching of an existing NLO-QCD calculation with parton-shower programs using the \POWHEG{} formalism and presented phenomenological results for a representative setup at the LHC. The code we developed is publicly available and can be adapted to the user's need in a straightforward manner. 

We have shown that theoretical uncertainties associated with the description of the third jet by  genuinely different parton-shower programs are mild at NLO+PS level, contrary to what is observed in studies based on matrix elements for VBF $Hjj$ production that are only LO accurate in the third jet. Our implementation thus provides an important improvement in the theoretical assessment of central-jet veto observables that are crucial for VBF analyses at the LHC.  
%
%
\section*{Acknowledgments} 
We are grateful to Carlo Oleari for help with implementing the code in the \POWHEGBOX{} repository. 
The work of
B.~J.\ is supported by the Institutional Strategy of the University of T\"ubingen (DFG, ZUK~63). 
F.~S.\ is supported by the ``Karlsruher Schule f\"ur Elementar\-teilchen- und Astroteilchenphysik: Wissenschaft und Technologie (KSETA)'' and D.~Z.\ by the BMBF under ``Verbundprojekt 05H2012 -- Theorie''.  

%

\end{document}